\documentclass[useAMS,usenatbib]{mnras}

\usepackage{float}
\usepackage{stfloats} % To put table* at the bottom
\usepackage{graphicx}
\usepackage{hyperref} % Allows creating Hyperlinks in Table of Contents and in-text References
\hypersetup{
    colorlinks,
    citecolor=blue, % Reference colour
    filecolor=black,
    linkcolor=blue, % ToC colour
    urlcolor=red
}

\usepackage{amsmath}
\usepackage{subcaption}
\usepackage{multirow}
\usepackage[flushleft]{threeparttable} % For table notes

\usepackage{siunitx}
\DeclareSIUnit\erg{erg}
\DeclareSIUnit\parsec{pc}

\title[Radio luminosity function at high-z]{Modelling the luminosities and sizes of radio sources: radio luminosity function at z = 6}
\author[A. Saxena et al.]{A. Saxena$^{1}$\thanks{E-mail:
saxena@strw.leidenuniv.nl}, H. J. A. R{\"o}ttgering$^{1}$ and E. E. Rigby$^{1}$\\
$^{1}$Leiden Observatory, Leiden University, P.O. Box 9513, 2300 RA Leiden, The Netherlands}
\begin{document}

% \date

\pagerange{\pageref{firstpage}--\pageref{lastpage}} \pubyear{2015}

\maketitle

\defcitealias{rig15}{R15}
\defcitealias{rig11}{R11}
\defcitealias{kai07}{KB07}

\label{firstpage}

\begin{abstract}
We present a model to predict the luminosity function for radio galaxies and their linear size distribution at any redshift. The model takes a black hole mass function and Eddington ratio distribution as input and tracks the evolution of radio sources, taking into account synchrotron, adiabatic and inverse Compton energy losses. We first test the model at $z=2$ where plenty of radio data is available and show that the radio luminosity function (RLF) is consistent with observations. We are able to reproduce the break in luminosity function that separates locally the FRI and FRII radio sources. Our prediction for linear size distribution at $z=2$ matches the observed distribution too. We then use our model to predict a RLF and linear size distribution at $z=6$, as this is the epoch when radio galaxies can be used as probes of reionisation. We demonstrate that higher inverse Compton losses lead to shorter source lifetimes and smaller sizes at high redshifts. The predicted sizes are consistent with the generally observed trend with redshift. We evolve the $z=2$ RLF based on observed quasar space densities at high redshifts, and show that our RLF prediction at $z=6$ is consistent. Finally, we predict the detection of 0.63, 0.092 and 0.0025 $z\ge6$ sources per sq. degree at flux density limits of 0.1, 0.5 and 3.5 mJy. We assess the trade-off between coverage area and depth and show that LOFAR surveys with flux density limits of 0.1 and 0.5 mJy would are the most efficient at detecting a large number of $z\ge6$ radio sources.
\end{abstract}


\section{Introduction}
Powerful high redshift radio galaxies (HzRGs) are found to reside in massive galaxies, which are thought to be progenitors of the massive ellipticals that we observe today \citep{bes98, mcl04}. These host galaxies contain huge amounts of dust and gas, and are observed to be forming stars intensively \citep{wil03}. HzRGs are also associated with cosmological over-densities such as galaxy clusters and proto-clusters \citep{rot03, ste03, kod07, ven07, gal12, may12}. These properties make HzRGs important tools to study the formation and evolution of massive galaxies and large-scale structure in the universe. For a review about the nature and properties of HzRGs, their hosts and their environments, we refer the reader to \citet{mil08}. 

Radio galaxies at the highest redshifts, particularly in the Epoch of Reionisation (EoR), have the potential to be important probes of cosmology. Constraining when and how the universe made a phase transition from neutral to completely ionised is one of the most exciting challenges in cosmology today and luminous radio galaxies at $z \ge 6$ may hold some clues on how this process unfolded. Evidence of the inter-galactic medium (IGM) being partly neutral at early times is found in the observed Gunn-Peterson trough \citep{gun65} in the spectra of $z \ge 6$ quasars \citep{bec01}. A robust signature of the neutral gas in the IGM could in principle be observed by measuring the hyperfine transition line of ground-state neutral hydrogen (with a rest-frame wavelength of 21 cm). At $z \ge 6$, this line falls in the low-frequency radio regime ($\nu < 200$ MHz) and may be detected as an absorption feature in the radio spectra of $z \ge 6$ radio galaxies and quasars \citep{car02, fur02}. Observing a 21-cm forest (similar to a Ly-$\alpha$ forest) in the continuum of a $z>6$ radio source can enable studying the process of reionisation over cosmic time \citep{xu09, mac12, ewa14, cia15}. Additionally, studying bright radio sources at $z \ge 6$ will further help understand the physics behind radiative processes responsible for ionising the universe.

As radio galaxies are often associated with some of the most massive supermassive black holes (SMBHs), HzRGs can be used to constrain black hole growth models. Understanding the growth of SMBHs over cosmic time using observations of active galactic nuclei (AGN) has been a subject of intense study. The growth of the black hole mass function has been studied using AGN and quasar luminosity functions and varying key parameters associated with AGN such as SMBH accretion rate, AGN duty cycles and radiative efficiencies \citep{sha09, kel10, sha13}. \citet{wil10b} managed to extend studies out to $z=6$. However, most observations used to study SMBH growth are at optical, infrared and Xray wavelengths. Not many AGN at high redshifts have been observed to be luminous at radio wavelengths and this could partly be due to a lack of deep, all-sky radio surveys. The general understanding, however, is that radio emission is powered by the same mechanism that is also responsible for the optical/IR and Xray emission from AGN, the accretion of material on to the central SMBH.

Many attempts have been made to measure the evolution of the radio luminosity function (RLF). \citet{dun90} reported evidence for the existence of a redshift cut-off in the space density of quasars and radio galaxies over the redshift range $2-4$. Subsequent studies also indicated that space densities of radio sources undergo continued decline between $z\simeq2.5$ and $4.5$ \citep{jar01}. This is similar to what optical studies of quasars have revealed, i.e. a peak in space density between redshifts 1.7 and 2.7 \citep{sch95} and a decline at higher redshifts \citep{fan04}. A decline in space density at $z>2.5$ was also reported in radio studies of flat spectrum quasars \citep{wal05}. Further, studies of X-ray selected AGN have shown that the space density of luminous AGN peaks between $z \sim 2$ \citep{has05} and $z \sim 2.5$ \citep{sil05}. More recently, \citet{rig11} used radio data available in the literature at the time to study the density evolution of steep-spectrum sources and found the redshift evolution of space densities to be dependent on luminosity. They found that the redshift at which space density of AGN with higher radio luminosities peaks, $z_{\textrm{peak}}$, is higher, thereby demonstrating that $z_{\textrm{peak}}$ is a function of radio luminosity. \citet{rig15} extended their earlier study to even lower radio luminosities and found consistent results.

It is largely agreed, however, that the RLF at any epoch is dominated by two distinct classes of objects -- star-forming systems at the lowest luminosities, which are generally hosted by late-type galaxies, and radio-loud AGN at higher luminosities, generally found to reside in massive early-type galaxies and powered by accretion on to the central SMBH \citep{jac99}. Radio-loud AGN can further be divided into two categories based on their radio morphologies \citep{fan74}. The Fanaroff-Riley class I (FRI) objects are brightest at their centres and have intermediate radio luminosities, whereas the FR class II (FRII) objects are brightest at the edges, away from the central regions and are some of the most luminous radio sources observed. \citet{dun90} showed that the dividing line between the FR classes, generally thought to be around $10^{25}$ W Hz$^{-1}$ at $150-200$ MHz, is remarkably close to the break in the local RLF. 

Observations have been complemented by theoretical studies aiming to understand the evolution of the radio luminosities and sizes of both FRI and FRII objects. These include modelling the dynamics of a radio source powered by relativistic jets resulting from the accretion on to the SMBH \citep{kai97a} and modelling the different physical processes through which a radio source may lose energy \citep{kai97b, blu99, ale02}. Some studies have also explored the link between the growth of the two FR classes \citep{ale00, kai07}. However, not a lot of work has gone into painting a complete picture that incorporates the growth of black holes with the evolution of individual radio sources.

In this work, we attempt to model the radio luminosity function based on black hole mass functions. Such an approach naturally establishes a link between the the growth of black holes and the resulting radio luminosity function, which can be tested using existing and upcoming radio surveys. We begin by testing our model at redshift 2, for which there are sufficient observations available. We then extend our model to $z=6$ with the ultimate goal of predicting a radio luminosity function close to or even into the EoR. Finally we use the modelled RLF to make predictions about the number of radio sources that could be observed in the current and upcoming state-of-the-art low-frequency radio surveys.

The layout of this paper is as follows. In Section \ref{sec:bhmf_model} we construct and describe our model for the growth and evolution of luminosities and linear sizes of radio sources. In Section \ref{sec:z2} we present luminosity and size predictions from our model at $z=2$ and test them using data available in the literature. We extend our model to $z=6$ in Section \ref{sec:z6} and present our predicted luminosity function and linear size distribution. We then test the predictions at $z=6$ by comparing with a RLF obtained using a pure density evolution model extrapolated from lower redshifts. We present the number of expected sources in current and future low frequency radio surveys and assess the trade-off between coverage area and depth, determining the optimum survey parameters to maximise detection of radio sources at high redshifts. Finally, we present a summary of our findings in Section \ref{sec:summary}.

Throughout this paper we assume a flat $\Lambda$CDM cosmology with $H_0 = 67.8$ km/s/Mpc and $\Omega_m = 0.307$. These parameters are taken from the first Planck cosmological data release \citep{pla14}.

\section{Modelling radio luminosities and linear sizes}
\label{sec:bhmf_model}

\subsection{Input black hole mass function and Eddington ratio distribution}
It is generally believed that radio galaxies are powered by accretion of material on to the supermassive black hole (SMBH). Therefore, we take black hole mass obeying the black hole mass functions (BHMF) determined at different epochs by \citet{sha09} as one of the inputs. These BHMFs have been derived using AGN bolometric luminosity functions estimated using optical and X-ray observations, making certain assumptions about radiative efficiency. The BHMF is generally well fit by a Schechter function of the form
\begin{equation}
	\phi(M_{BH}) = \phi_{\star} \left(\frac{M_{BH}}{M_{\star}}\right)^\alpha \exp \left[-\frac{M_{BH}}{M_{\star}}\right]
\end{equation} 
where $\phi_{\star}$ and $M_{\star}$ are the characteristic space density and black hole mass, respectively and $\alpha$ is the low-mass end slope.

For an actively accreting SMBH with an accretion rate $dM/dt$, the bolometric luminosity can be written as $L_{bol} = \epsilon (dM/dt) c^2$, where $\epsilon$ is the efficiency parameter and $c$ is the speed of light. The maximum possible luminosity achievable through this mechanism is the Eddington luminosity, $L_{Edd}$ and the ratio of the bolometric luminosity to the Eddington luminosity, the Eddington ratio, is written as $\lambda = L_{bol}/L_{Edd}$. This can take values between 0 and 1, and is an indicator of how `actively' a SMBH is accreting. The Eddington ratio is another input in our model, which is drawn from a log-normal distribution that \citet{sha13} found to fit the observed AGN luminosity functions well and is also supported by \citet{wil10b} at $z \sim 6$.

\subsection{Radio jet power calculation}
The jet power is thought to be closely coupled to the black hole mass, spin and accretion rate via the Blandford-Znajek mechanism \citep{bla77}. Jet power can be calculated either by using the thin disk solution \citep{sha73}, which typically works for black holes accreting at higher Eddington ratios ($\lambda > 0.01$) or by assuming a thick accretion disk with an advection dominated accretion flow (ADAF) for black holes accreting at lower rates \citep{nar94}. The expression for jet power in the thin disk regime can be written as \citep{mei02, ors16}
\begin{equation}
	Q_{\textrm{jet}} = 2 \times 10^{36} \left(\frac{M_{BH}}{10^9 M_{\sun}}\right)^{1.1} \left(\frac{\lambda}{0.01}\right)^{1.2} a^2 \hspace{2mm} \si{\watt}
	\label{eq:jetpower_thin}
\end{equation} 
and in the ADAF regime can be written as
\begin{equation}
	Q_{\textrm{jet}} = 2.5 \times 10^{38} \left(\frac{M_{BH}}{10^9 M_{\sun}}\right) \left(\frac{\lambda}{0.01}\right)  a^2 \hspace{2mm} \si{\watt}
	\label{eq:jetpower_thick}
\end{equation}
where $M_{BH}$ is the black hole mass, $\lambda$ is the Eddington ratio and $a$ is the black hole spin. We use a Monte Carlo based approach where a value of $\lambda$ is randomly sampled from the given input distribution (which we elaborate upon in the next section) and assigned to a black hole mass to calculate jet powers for each source using Equations \ref{eq:jetpower_thin} or \ref{eq:jetpower_thick}, depending on the Eddington ratio. 

Although studies have indicated that AGN might experience recurrent jet activity through refuelling of the central black hole due to a possible merger, the lifetime of a luminous radio source is typically of the order of $10^{7}$ yr, with the AGN duty cycle found to be close to $10^9$ yr \citep{bir10}. Note that duty cycle values have been shown to strongly depend on the stellar mass of host galaxies \citep{bes05}. However, the typical duty cycle timescale is longer than the time for which we evolve our sources, which we talk about in the following sections. Therefore, any recurrent jet activity is not taken into account in our modelling. 

\subsection{Radio luminosity calculation}
Several studies have aimed to establish an empirically derived relation between the intrinsic jet power and the observed radio luminosity \citep{bir04, bir08, cav10}. There have also been several attempts to construct analytical models for the growth of jets in radio sources, especially in strong radio sources with an FRII-type morphology that are typically associated with HzRGs \citep{kai97a, kai97b, blu99, ale00, ale02}.

\citet[][hereafter \citetalias{kai07}]{kai07} studied analytically the growth of radio galaxies in the context of the radio luminosity function and determined that the luminosity function at any epoch consists of sources that are dominated by different energy loss mechanisms, depending on their ages and sizes. Each energy loss phase affects their growth and the evolution of their radio luminosity differently. We use the \citetalias{kai07} prescriptions to track the growth of radio sources in our model, which contribute to the luminosity functions. We briefly describe the implemented energy loss mechanisms below.

\subsubsection{Synchrotron losses}
Synchrotron radiation from ultra-relativistic charged particles is believed to be the major source of virtually all extragalactic radio sources. In star-forming galaxies, synchrotron radiation originates from electrons from HII regions, accelerated by Type II and Ib supernovae. In radio-loud AGN, charged particles are accelerated in the jets and lobes launched by the central SMBH due to accretion of matter, and such systems are the focus of this model.

The energy density of the magnetic field associated with a radio source decreases as a source grows. Therefore, synchrotron emission dominates early in the source's lifetime. \citetalias{kai07} noted that the luminosity of a source growing in this regime stays constant, with the luminosity at an observing frequency $\nu$ given by
\begin{equation}
	L_\nu \approx \frac{m_e c^2 f_n}{6 A^{1/2} \nu} Q_{jet}
	\label{eq:synchrotron}
\end{equation}
where $m_e$ is the mass of an electron, $c$ is the speed of light, $f_n$ and $A$ are constants with fiducial values $1.2 \times 10^{12}$ s$^2$kg$^{-1}$m$^{-2}$ and $4$, respectively.

\subsubsection{Adiabatic losses}
After the early stages of synchrotron losses, the magnetic field strength declines and adiabatic losses begin to dominate. The gas density around a radio-loud AGN into which the jets and lobes grow is generally represented by a single-$\beta$ model of the form
\begin{equation}
	\rho_{r} = \frac{\rho}{[1+(r/d)^2]^{\beta/2}}
	\label{eq:gasdensity}
\end{equation}
where $\rho$ is the density in the inner-most regions, $d$ is the distance until which the gas density remains constant (and follows a power-law decline afterwards) and $r$ is the distance from the centre of the gas distribution. Such a profile has been found to fit X-ray emission from hot gas in elliptical galaxies (that usually host radio-loud AGN), galaxy groups and clusters \citep{fuk04}.

When the source is in the adiabatic loss phase, the luminosity evolves as \citepalias{kai07}
\begin{equation}
	L_{\nu} \propto D^{(8-7\beta)/12}
\end{equation}
where $D$ is the linear size of the radio lobe and $\beta$ depends on the profile of the ambient gas surrounding the radio source.

\subsubsection{Inverse Compton losses}
Once the size of the lobe has exceeded the extent of the x-ray halo of the host galaxy, and the energy density of the magnetic field in the lobe is comparable to the energy density of the ambient cosmic microwave background (CMB) radiation, losses due to inverse Compton scattering against CMB photons begin to dominate. \citetalias{kai07} found that the luminosity in this phase evolves as
\begin{equation}
	L_{\nu} \propto D^{(-4-\beta)/3}
\end{equation}

The CMB energy density scales with redshift, $z$, as $\propto(1+z)^4$. Therefore at higher redshifts, IC losses should begin to dominate much earlier in the source's lifetime. We calculate the distance from the centre of the galaxy after which IC losses begin to dominate, $D_{IC}$, by equating the luminosities produced by adiabatic losses and IC losses, given by equations (3) and (7) in \citetalias{kai07}. This depends on the jet power of the source and can be written as (See Appendix \ref{app:dic})
\begin{equation}
	D_{IC} \propto (\rho d^\beta)^{-1/6} Q_{jet}^{1/3} (1+z)^{-8/3}
\end{equation}	
	
\subsubsection{Overall evolution and RLF determination}
Having set up the various phases of evolution, we can track the luminosity and size evolution of all our sources. Equation (A2) in \citetalias{kai07} describes the growth of the lobe size, D with time
\begin{equation}
	D = C \left(\frac{Q_{jet}}{\rho d^\beta}\right)^{1/(5-\beta)} t^{3/(5-\beta)}
\end{equation}
where $C$ is a constant. For simplicity, we use the fiducial values from \citetalias{kai07} and set the extent till which the gas density remains constant to $d = 2$ kpc at $z=2$. We then include redshift evolution of linear sizes of galaxies. \citet[][and references therein]{wel14} find that over the redshift range $0<z<3$, the size evolution of early-type galaxies is much faster than late-type galaxies. Further, \citet{mos12} find that the sizes of Lyman break selected galaxies (LBGs) with stellar masses of $10^{9.5} < M_{\star} < 10^{10.4}$ evolve as $(1+z)^{-1.2}$ from $z=1$ to 7, which is consistent with the findings of \citet{wel14}. At higher redshifts, $7<z<12$, \citet{ono13} find the size evolution of LBGs to follow $(1+z)^{-1.3}$. Radio galaxies are more likely to be hosted by early-type galaxies up to redshifts of $z=3$, but at higher redshifts, HzRGs are seen to be forming stars intensively \citep{mil08}.

Owing to the faster evolution of early-type galaxies seen at moderate redshifts and the redshift evolution of LBGs at higher redshift, and assuming that early in the universe radio galaxies are expected to be hosted by galaxies with properties similar to LBGs, we include the size evolution with redshift as $(1+z)^{-1.25}$, which is consistent with the majority of studies carried out at $z>2$. Therefore, the parameter $d$ takes values
\[
	d = 
    \begin{cases}
    	2 \text{ kpc},& \text{if } z \le 2 \\
        2\times[(1+z)/3]^{-1.25} \text{ kpc}, & \text{if } z > 2
    \end{cases}
\]
The gas density in the inner parts of the galaxy is set as $\rho = 10^{-22}$ kg m$^{-3}$ at all redshifts.

The initial growth of the lobe would be in a region where the ambient gas density is roughly constant, i.e. $D < d$ and therefore, $\beta=0$. This very early growth would be dominated by synchrotron losses as the magnetic field is influential.  Considering Equation \ref{eq:synchrotron}, the luminosity would level off at the stage when synchrotron radiation dominates. After the lobe has grown to a size of 2 kpc, the source enters the adiabatic loss phase. Here, we consider the gas density profile around the radio source to follow a power law decline (Equation \ref{eq:gasdensity}), with $\beta=2$ being used for further evolution. Finally, once the source has grown to a size when the CMB energy density begins to play an important role, inverse Compton (IC) losses begin to dominate. The values of all model parameters used in this study are shown in Table \ref{tab:params} in Appendix \ref{app:dic}.

\section{Radio luminosities and linear sizes at z = 2}
\label{sec:z2}
\subsection{Model Input}
\begin{figure}
	\centering
	\includegraphics[scale=0.45]{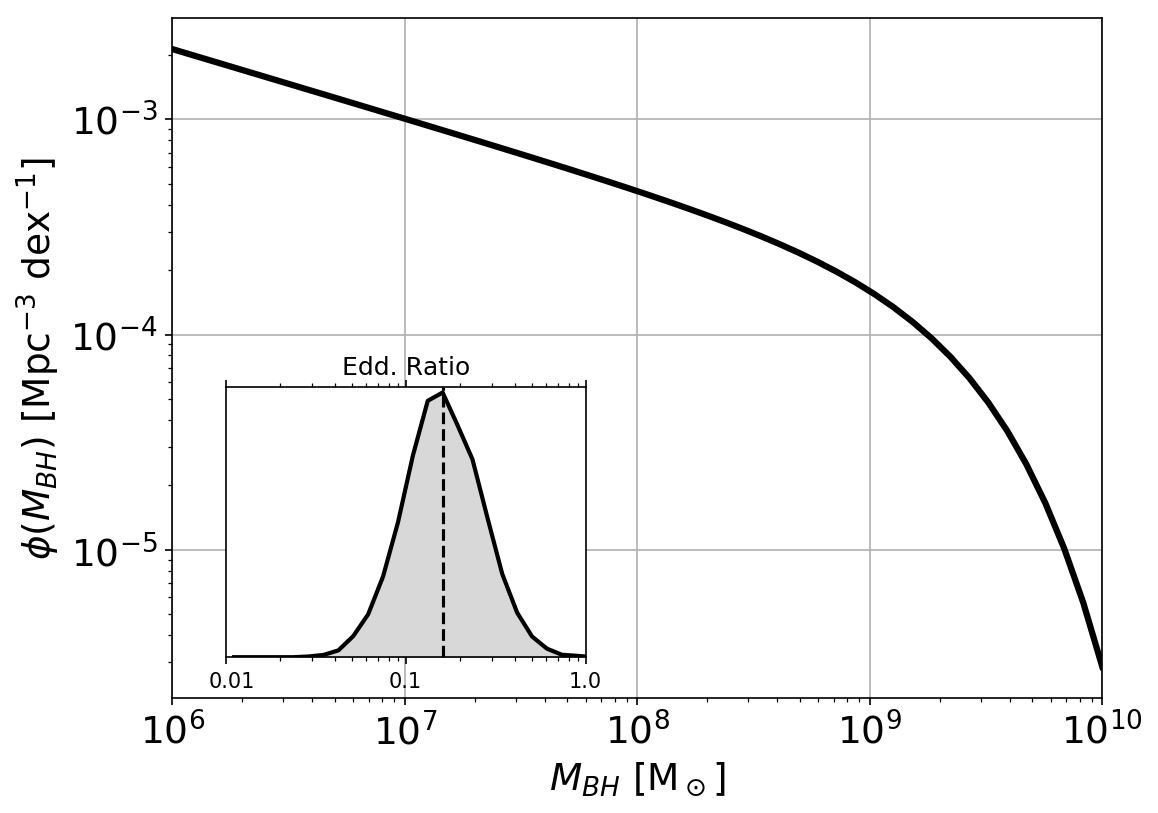}
	\caption{Input black hole mass function and the Eddington ratio at $z=2$ taken from \citet{sha09} and \citet{sha13}. Shown in the figure is the best fit Schechter function determined at $z=2$. The Eddington ratio distribution is a log-normal peaking at $\lambda=0.16$.}
	\label{fig:bhmf_z2}
\end{figure}
We first implement our model at $z=2$ to compare the predictions of our model to data, as the radio luminosity function at this epoch is relatively well constrained \citep{dun90, blu99, wil01, rig11, rig15}. We use the \citet{sha09} BHMF determined at $z=2$, which is a Schechter function with parameters $\log \phi_\star = -3.665$, $\log M_\star = 9.076$, and $\alpha=-0.3223$, and the lognormal Eddington ratio distribution peaking at $\lambda = 0.16$ with dispersion of 0.5 dex from \citet{sha13} (both shown in Figure \ref{fig:bhmf_z2}) as input. For simplicity, we assume the Eddington ratio to be independent of the black hole mass. We randomly sample a value of the Eddington ratio from the chosen distribution and assign it to each black hole. 

We assume a mass dependence for the spin parameter of our black holes, following the conclusions of \citet{vol07}. They note that disk galaxies harbour lower mass SMBHs and weaker AGN, and grow by accreting smaller packets of material. This would skew the black hole spin distribution to lower values. Brighter radio sources, however, are found in elliptical galaxies that host massive SMBHs. These black holes must have had a major accretion episode, likely powered by a merger that was responsible for forming the host elliptical galaxies too. During this episode the spin must have increased significantly. \citet{vol07} show that the peak in distribution of spin parameter for black holes with masses greater than $10^8$ $M_\odot$ lies between $0.7-1.0$. Therefore, black holes with masses greater than $10^8$ $M_\odot$ in our simulation are randomly assigned a spin from the range $[0.7, 1.0)$ and all other black holes are assigned a spin from the range $[0.0, 0.7)$. As a test, we implement two other spin distribution schemes, one where the mass dependence is reversed, i.e. the smaller mass black holes have higher spins, and the other where all black holes have a constant spin of 0.6. The black hole mass, Eddington ratio and spin parameter are the required inputs to calculate jet powers for all sources in our simulation.

\subsection{Output jet power distribution}
\begin{figure}
	\centering
	\includegraphics[scale=0.45]{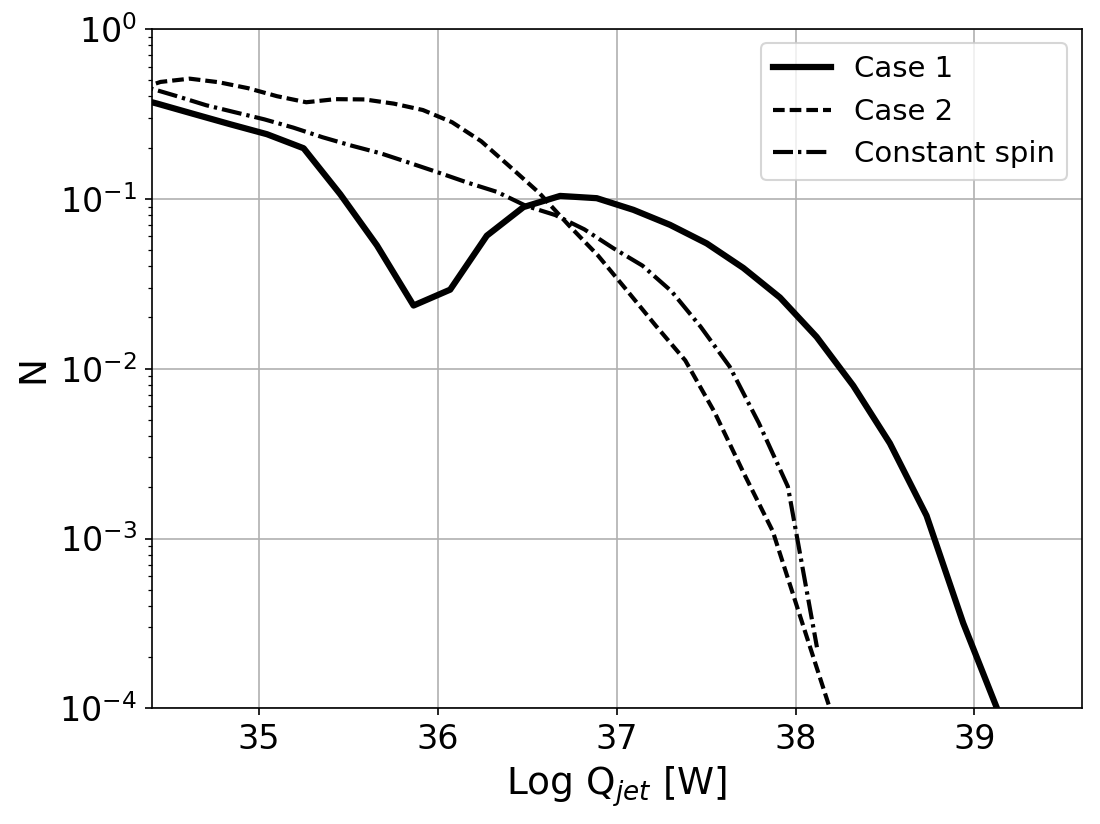}
	\caption{Distribution of jet powers predicted by our model at $z=2$ for different choices of spin parameters. Case 1 is when more massive black holes have a higher spin, Case 2 is when more massive black holes have lower spin and Case 3 is when all black holes have a constant spin. Case 1 most accurately reproduces the bimodality in jet power, and is able to produce sources with very powerful jets that result in some of the largest and brightest radio galaxies in the universe.}
	\label{fig:z2_jetpower}
\end{figure}

The resulting jet power distributions for the 3 different choices of spin parameter distributions are shown in Figure \ref{fig:z2_jetpower}. Cases where smaller black holes have higher spin (Case 2) and a constant spin parameter (Case 3) are unable to produce the most powerful jets, which have been observed in the literature \citep[see][]{god13}. The widely supported idea that more massive black holes are highly spun up (Case 1) accurately reproduces the bimodal distribution of jet powers, which is thought to result in the fundamental difference between the FRI and FRII type radio galaxies. This bimodality is likely due to the two different accretion regimes implemented in our model (ADAF and thin disk regimes), that depend on the accretion rate. Therefore, going forward, we only consider results from Case 1 of spin parameter choice.

\subsection{Constructing and normalising the radio luminosity function}
The first step at which we construct the luminosity function is when the linear size of all sources in the simulation is 2 kpc, i.e. when all sources are dominated by synchrotron losses. Since the growth rate depends on jet power, the radio sources take different times to attain a size of 2 kpc, thus leading to an age distribution. To now track the evolution of radio luminosities, we evolve our sources in time steps of 0.2 Myr, calculating linear sizes at each time step. The size determines which phase of energy loss each source is in and we use this to calculate the evolved luminosities at each time step according to the prescriptions described in the previous section. The simulation is run for a total time of 9 Myr, which is the typical lifetime of a radio galaxy.

Although it has been shown that the radio-loud fraction of AGN is a function of the stellar mass of the host galaxy and the black hole mass in the local universe, these values are relatively unconstrained at high redshifts \citep{bes05,wil15}. The physical conditions of the universe change dramatically going from $z=0$ to $z=2$, so a simple extrapolation from the local universe would not work. It has been observed however, that roughly 10\% of all galaxies are AGN \citep[][and references therein]{mar13} and of these, only 10\% are generally found to be bright in the radio or radio-loud, even out to higher redshifts \citep{ban15}. Further, the correlation between stellar mass and radio-loud fraction is weaker for high-excitation radio galaxies \citep{jan12}, which most radio sources at high redshifts are expected to be.  Therefore, we take a simplistic approach and randomly select $1\%$ of all our sources to be included in the final radio luminosity function. It is important to note that we do not select objects to be AGN or radio-loud depending on the black hole mass, as black hole masses of (radio-loud) AGN seem to be distributed evenly over several orders of magnitude \citep{woo02}. 

We use the \citet{sha09} black hole space densities to normalise the radio luminosity function in the following way. The space densities from the analytical BHMF are used to assign a maximum volume, $V_{\textrm{max}}$, to each simulated black hole, which is the volume probed by a complete survey that would enable the black hole to be detected. These calculated volumes are then used to normalise the resulting radio luminosity function, by summing over $1/V_{\textrm{max}}$ in each luminosity bin, which is the traditional way of calculating luminosity functions \citep{sch68}.
\begin{figure}
	\centering
	\includegraphics[scale=0.45]{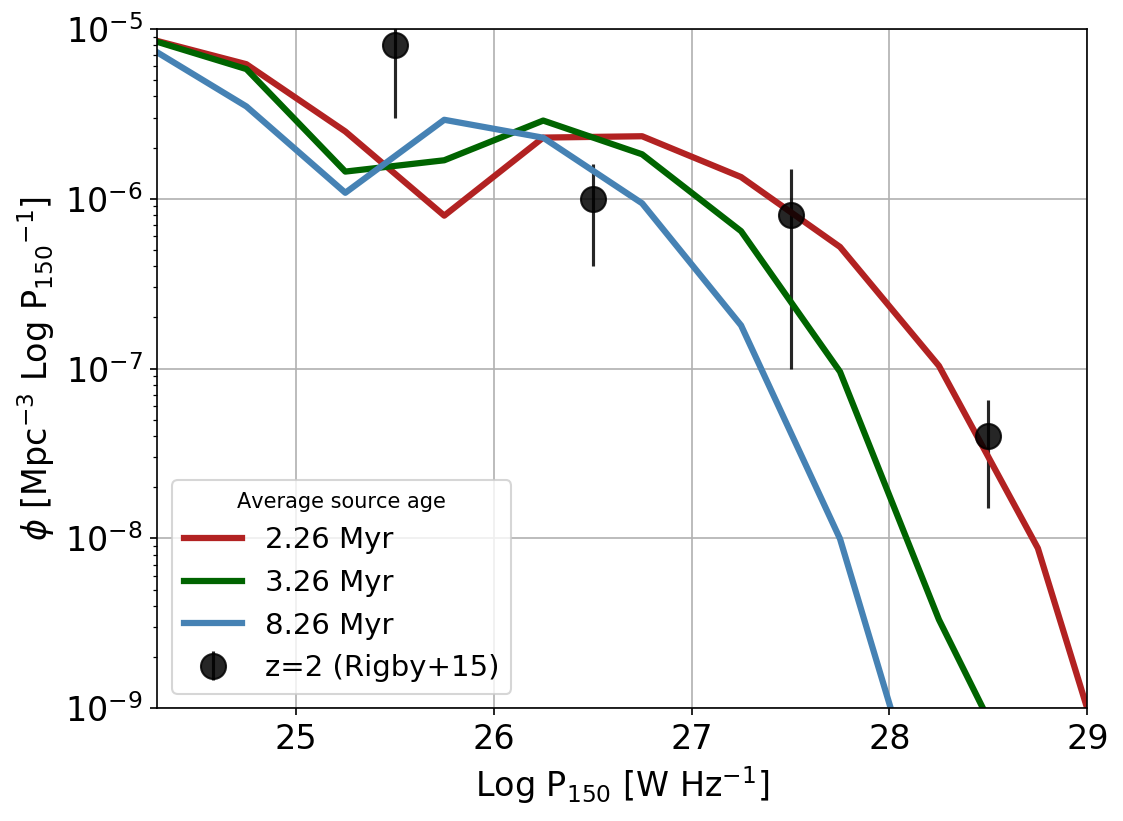}
	\caption{Time evolution of the radio luminosity function (RLF) at $z=2$ at various time steps in our model. Each curve shows the predicted luminosity function at a given average source age. The brightest sources lose their energy very quickly and therefore, those observed must be young. Also shown are the space densities calculated by \citet{rig15}, which have been recalculated at 150 MHz. Additionally, two distinct radio populations are apparent, with the dividing luminosity lying between $10^{25}$ and $10^{26}$ W Hz$^{-1}$, which is consistent with the luminosity that is generally considered as the dividing line between the FRI and FRII sources \citep{dun90, wil01}.}
	\label{fig:rlf_z2}
\end{figure}

\subsection{Output luminosity functions}
Luminosity distributions constructed for 3 time steps are shown in Figure \ref{fig:rlf_z2}. The bright end of the luminosity distribution is seen to change, whereas the faint end remains roughly constant over time. This is mainly because the most powerful sources in the simulation grow faster and this rapid growth leads to increased adiabatic losses. Additionally, sources with powerful jets enter the regime of Inverse Compton losses and end up losing energy much quicker. This suggests that the most powerful sources must be very young and as a result, compact.

We plot the space densities calculated by \citet{rig15} for comparison. Since the space densities were calculated at an observing frequency of 1.4 GHz, we scale it to obtain powers at 150 MHz using the $z-\alpha$ relation $\alpha = 0.8 + 0.21\log (1+z)$, determined by \citet{ker12}. Overall, the data seems to match the predictions well and is also consistent with the expectation that most powerful sources must be younger. There is a slight disagreement in the luminosity bin $25 < \log P < 26$, where the observed space densities are higher than our prediction. However, this may be explained by the apparent presence of two distinct populations, with the dividing luminosity between $10^{25} - 10^{26}$ W Hz$^{-1}$. This coincides with the dividing luminosity between FRI and FRII radio sources, which is seen in observations \citep{dun90, wil01} and was also predicted by \citetalias{kai07}.

Another interesting feature especially prominent around this dividing luminosity, is the transition of sources from the FRII to the FRI regime over time (purely on the basis of radio power; within the framework of this simulation, we are unable to separate the sources morphologically). Such a scenario was suggested by \citetalias{kai07}, where the development of an FRII structure in a source with a weak jet can be disrupted by dense external gas, which starts to fill the lobe volume, eventually reaching the jet flow. At this point the morphology is expected to change from FRII to FRI accompanied by a decrease in radio luminosity, which is what we may be seeing in our simulation. Such a behaviour, combined with the apparent bimodality in the distribution of jet powers as seen in Figure \ref{fig:z2_jetpower} suggests that the origin of the FRI and FRII populations could be purely due to the mechanism by which the central SMBH is accreting gas. Such a scenario was suggested by \citet{bes12}, where they constructed a large sample of radio sources from the SDSS and classified them broadly into two categories based on their emission line properties, low-excitation radio galaxies (LERGs) and high-excitation radio galaxies (HERGs). They found LERGs to dominate at low radio luminosities and HERGs to dominate at high luminosities, with the switch in population dominance found to be at $L_{\textrm{1.4 GHz}} = 10^{26}$ W Hz$^{-1}$. LERGs are powered by radiatively inefficient accretion, with much lower accretion rate compared to the radiatively efficient accretion that powers HERGs. This divide in luminosity is roughly where we observe two separate populations too, which we attribute to the difference in accretion mechanism. Therefore, the scenarios suggested by \citet{bes12} and what we see in our simulation seem to be consistent.

\subsection{Output size distribution}
\begin{figure}
	\centering
	\includegraphics[scale=0.45]{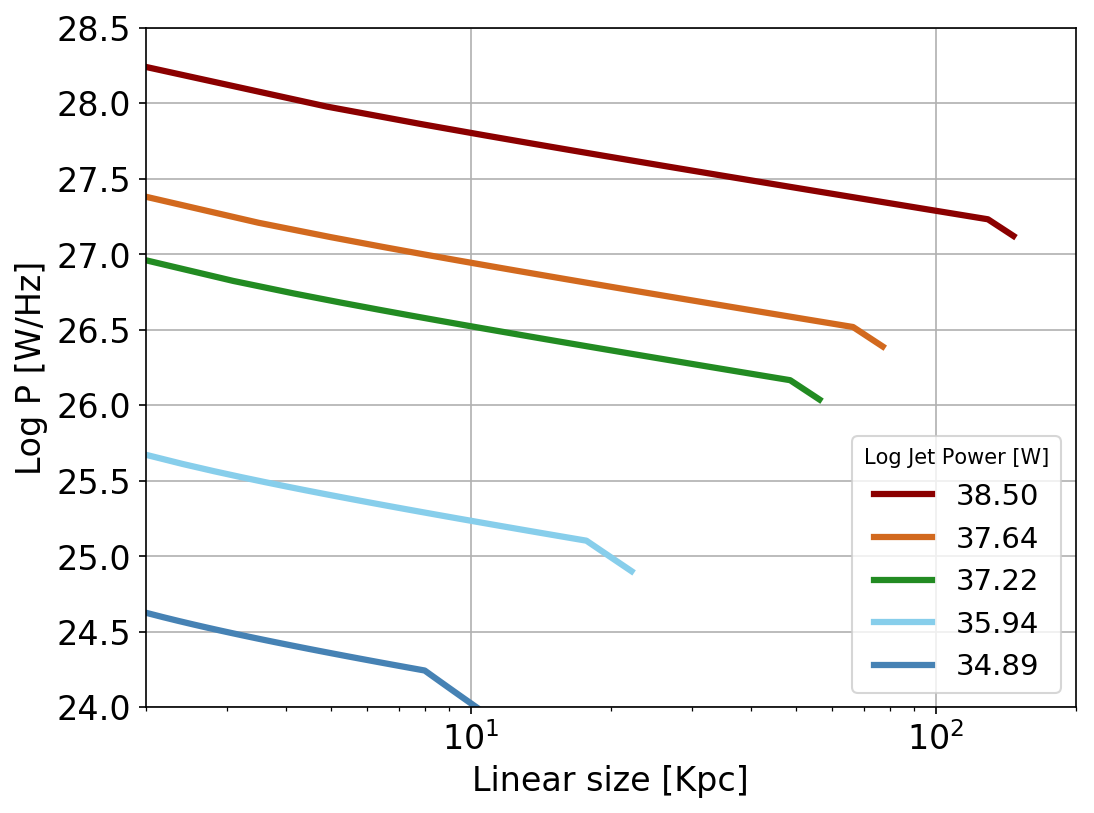}
	\caption{The so-called P-D tracks, showing the evolution of radio power (P) with linear size (D) for sources with different radio jet powers at $z=2$. Sources with stronger jets have a well behaved evolution, whereas sources with weaker jets have a break in their track very early in their lifetime. This is because weaker sources enter the regime of inverse Compton losses much quicker.}
	\label{fig:pdtracks}
\end{figure}

The included energy-loss prescriptions in our model give rise to `P-D tracks' for every source in our simulation, which have been historically used to study the evolution of radio lobes in FRII galaxies \citep{blu99, ale00, ale02, kai07}. Figure \ref{fig:pdtracks} shows P-D tracks for sources with different jet powers for the entirety of the simulation run (9 Myrs). Sources with low jet powers show a break in their P-D track, which arises when there is a decline in their radio luminosities after attaining a certain linear size, when the dominant energy loss mechanism changes from adiabatic losses to inverse Compton (IC) losses from the CMB. Growth of the source in the IC regime is slower. Sources with high jet powers follow, however, are more likely never to be dominated by IC losses and follow a well-behaved path along the P-D diagram.

Since the average linear sizes of observable sources in our model depend strongly on the average age of sources at a given time step, we weigh the observed sizes by age. This is done by assigning a probability of detection, which is defined as the age of a source divided by its total lifetime in the simulation. This means that younger sources have a lower probability of detection than sources that have existed in the universe for a longer time. We construct a linear size distribution in time steps of 1 Myr and normalise this distribution by assigning a detection probability to each source. 
\begin{figure}
	\centering
	\includegraphics[scale=0.45]{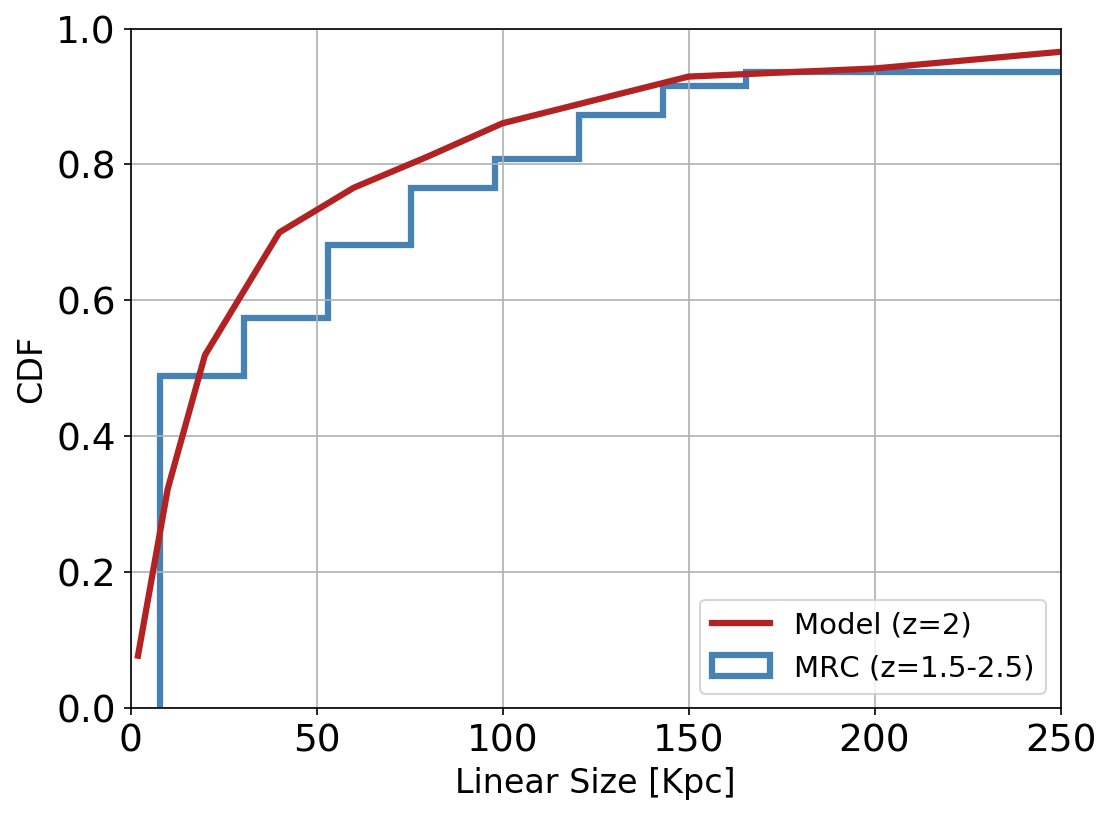}
	\caption{Comparison of the cumulative distribution of linear sizes of FRII radio sources predicted by our model (red line) with radio galaxies with confirmed redshifts in the range $z=1.5-2.5$ from the Molonglo Reference Catalogue (MRC; blue line). The distributions seem to agree well, with our model not under- or over-predicting very small or very large linear sizes. The MRC contains 3 sources with sizes greater than 250 kpc in the selected redshift range and since our model at $z=2$ is unable to produce sources with sizes greater than 250 kpc, we have excluded these particular MRC sources in this comparison for clarity.}
	\label{fig:sizes_mrc}
\end{figure}

To test the size predictions from our model, we use the catalogue compiled by \citet{sin13}, which contains linear sizes and redshifts for radio sources in the Molonglo Reference Catalogue (MRC; \citealt{lar81}). The original MRC contains 12,141 discrete sources with flux density $\ge0.7$ Jy and the survey covers 7.85 sr. of the sky. For this comparison we only consider sources in the redshift range $1.5-2.5$ in the \citet{sin13} catalogue. We also implement a flux cut in our predicted observations that matches the completeness level of the sample. The cumulative distribution functions of linear sizes predicted by our model and the MRC are shown in Figure \ref{fig:sizes_mrc}. The size distributions seem to agree and our model does not over- or under-predict very large or small sources. Within the selected redshift range however, the \citet{sin13} sample contains 3 sources with linear sizes $>250$ kpc (263, 284 and 458 kpc at redshifts 1.54, 1.78 and 1.54 respectively). Our model at $z=2$ is unable to produce sources with sizes greater than 250 kpc. This could be because we do not model recurrent jet activity, which may be responsible for the continued growth of a radio source by periodically refuelling the jet from the SMBH. Therefore, in the comparison shown in Figure \ref{fig:sizes_mrc}, we have excluded the three largest sources from the MRC. It is safe to assume, however, that the inclusion of the largest sources should not affect the cumulative distribution of linear sizes too much as these sources lie well above the mean linear size.

\section{Implementing the model at z = 6}
\label{sec:z6}
Having tested the predictions of our model at $z=2$ where there are sufficient observations available, we now extend the model to $z=6$ to predict a radio luminosity function at this redshift, and describe the implementation in the following section.

\subsection{Model input}
We use the black hole mass function (BHMF) and Eddington ratio distribution at $z=6$ determined by \citet{wil10b} using the quasar luminosity function and estimation of black hole masses through measurements of MgII line widths. The luminosity function they use to derive the BHMF has been found to be consistent with more recent studies of $z\sim6$ quasars exploring both fainter and brighter populations \citep{kas15,jia16} The best-fit Schechter function parameters for their BHMF assuming a duty cycle of 0.75 are $M_\star = 2.24\times 10^{9}$ $M_\odot$, $\phi_\star(M_{BH}) = 1.23\times 10^{-8}$ Mpc$^{-3}$ dex$^{-1}$ and faint-end slope $\alpha = -1.03$. Further, \citet{wil10b} found the Eddington ratio distribution at $z=6$ to be a log-normal distribution, peaking at 0.6 and with a dispersion of 0.30 dex, which is consistent with the distribution used by \citet{sha13} and this is what we use in our model.

We also account for the redshift-evolution of linear sizes of galaxies and how that affects the parameter $d$, which represents the extent of constant gas density around a source. This redshift evolution goes as $\propto (1+z)^{-1.25}$ as discussed in Section 2.3.4. Finally, in order for the black hole masses to be high enough at $z=6$, they must have been accreting very close to the Eddington limit from very early times after their formation and must be nearly maximally spun up \citep{sha05, dal09, dub14}. Therefore, we set the spin parameter at this redshift to $a=1$. The simulation is run for a total of 5 Myrs.

\subsection{Model predictions at z = 6}

\subsubsection{Radio luminosity function}
\begin{figure}
	\centering
	\includegraphics[scale=0.45]{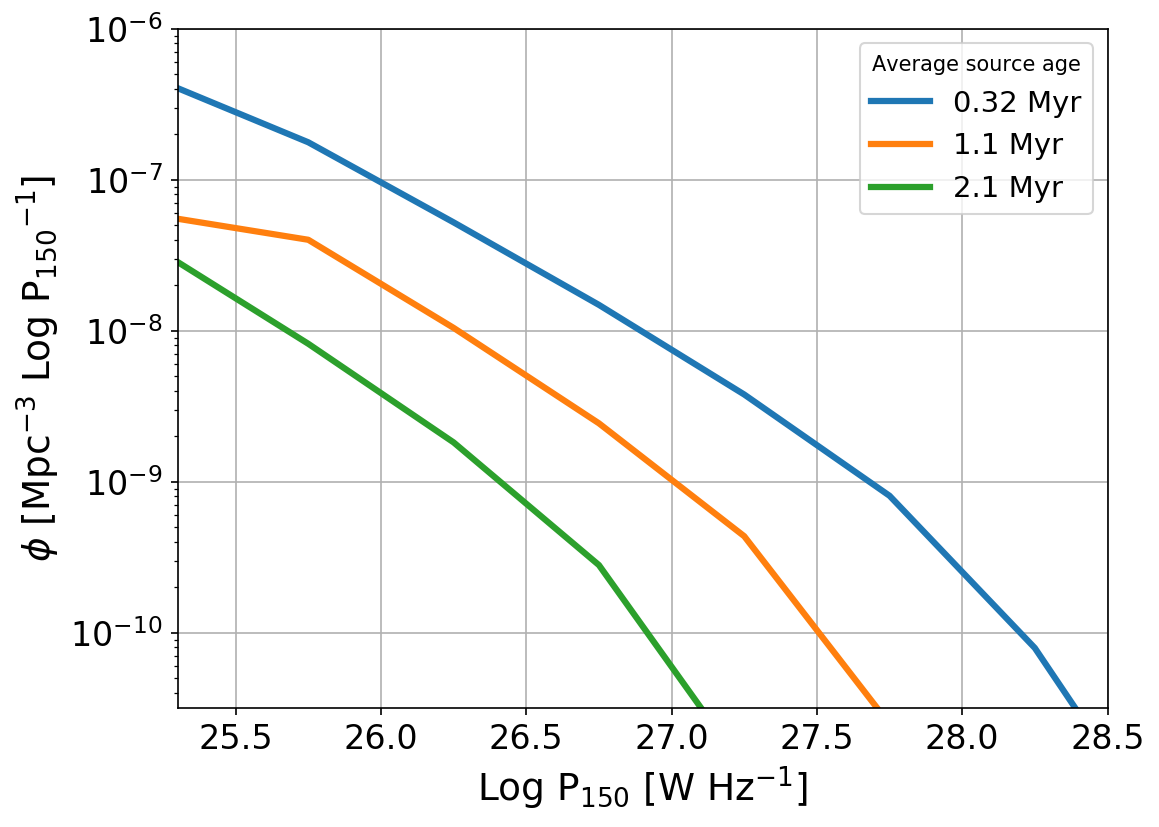}
	\caption{Luminosity functions at $z = 6$ at various stages of source evolution. The mean ages of the sources contributing to the luminosity function are shown. At this redshift, it is clear that sources are much younger and lose their energy much more rapidly. This increased loss of energy can be attributed to the much stronger CMB energy density, which increases the inverse Compton losses. After $\sim2$ Myrs, there are hardly any luminous sources left in the simulation. This suggests that luminous radio sources observed at $z=6$ must be very young. The faint end is seen to evolve less than the bright end.}
	\label{fig:rlf_mcgrowth}
\end{figure}
\begin{figure}
	\centering
	\includegraphics[scale=0.45]{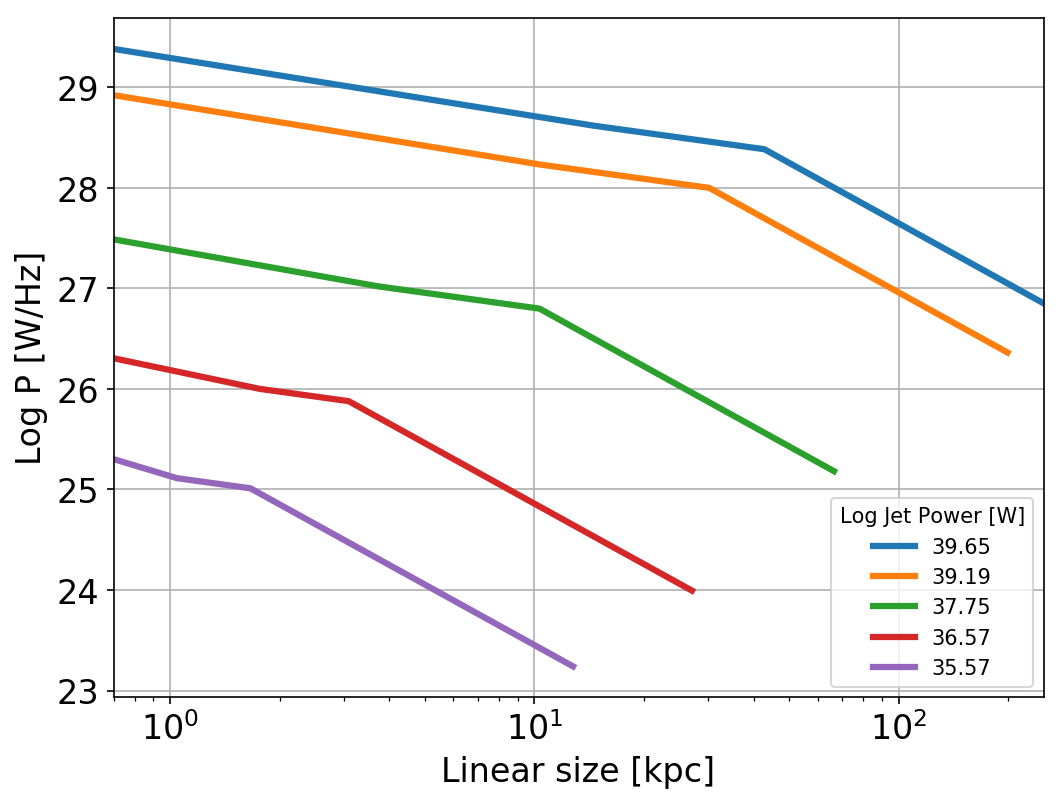}
	\caption{Comparison of the evolution of radio power with linear size (P-D tracks) for sources with given jet powers in our simulation at $z=6$. Clearly, the Inverse Compton losses come in to effect much quicker at higher redshifts, as evident from the break in the P-D tracks at much smaller linear sizes. Overall, sources at $z=6$ find it much harder to grow to large sizes due to the increased CMB energy density.}
	\label{fig:pd_z6}
\end{figure}

The time evolution of the distribution of radio luminosities at $z=6$ is shown in Figure \ref{fig:rlf_mcgrowth}. Note that only sources with FRII-like radio powers are shown, as less luminous sources at $z=6$ are beyond the detection capabilities of current radio surveys. There are only a handful of sources with very powerful radio luminosities ($> 10^{28}$ W Hz$^{-1}$). This is mainly due to the generally lower SMBH masses at higher redshifts and the increased inverse Compton losses due to the denser CMB. The time evolution of the luminosity function is much stronger too, which can be attributed to rapid energy losses. This is especially prevalent at the most luminous end of the luminosity function, which can be seen from the `P-D' tracks for sources in our simulation, shown in Figure \ref{fig:pd_z6}. More luminous sources are fuelled by powerful jets and grow faster. As a result, they enter the regime of strong Inverse Compton (IC) losses much earlier in their lifetime. At lower luminosities, evolution is slower as it consists of sources with relatively weaker jets that grow at a lower rate and lose their energy less rapidly. 

Overall at $z=6$, sources are younger and lose energy at a rate that is much higher than seen at lower redshifts. This has implications on the nature of sources that we can expect to observe at $z\ge6$. Any luminous source observed at this epoch must be very young and compact, consistent with the `inevitable youthfulness' of radio sources in the early universe, as suggested by \citet{blu99b}.

\subsubsection{Linear sizes}
\begin{figure}
	\centering
	\includegraphics[scale=0.45]{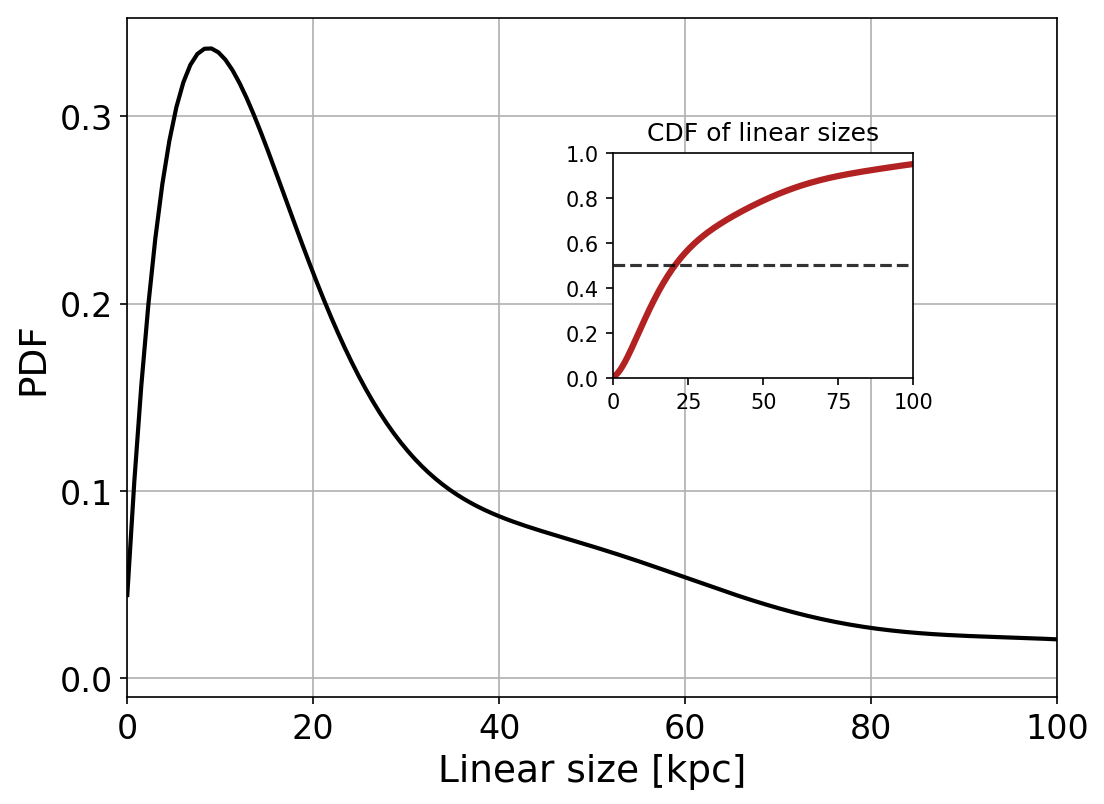}
	\caption{The age-weighted probability distribution function and the cumulative distribution (inset) of predicted linear sizes of radio galaxies at $z=6$. The mean linear size is expected to be roughly 25-30 kpc. Large sources at this epoch die out quicker due to the increased energy density of the CMB. Therefore, radio sources at $z=6$ are generally expected to be young and very compact.}
	\label{fig:sizes_z6}
\end{figure}

The higher CMB energy density prevents lobes from growing to large sizes (a few hundred kpc) and as a result, linear sizes at higher redshifts are generally smaller compared to lower redshifts. We apply the age-weighted normalisation introduced in Section 3.5, keeping in mind that the detection probability of younger sources is less than older ones. The resulting linear size distribution for sources brighter than 0.5 mJy at 150 MHz, which represents the lowest flux densities for the LOFAR Tier 1 survey \citep{shi17}, is shown in Figure \ref{fig:sizes_z6}. Also shown in the inset is the cumulative distribution function of the linear size distribution. We find that the median size of radio galaxies at $z=6$ is $20$ kpc, which translates to an angular scale of roughly $3.5$ arcseconds on the sky.

\begin{figure}
	\centering
	\includegraphics[scale=0.45]{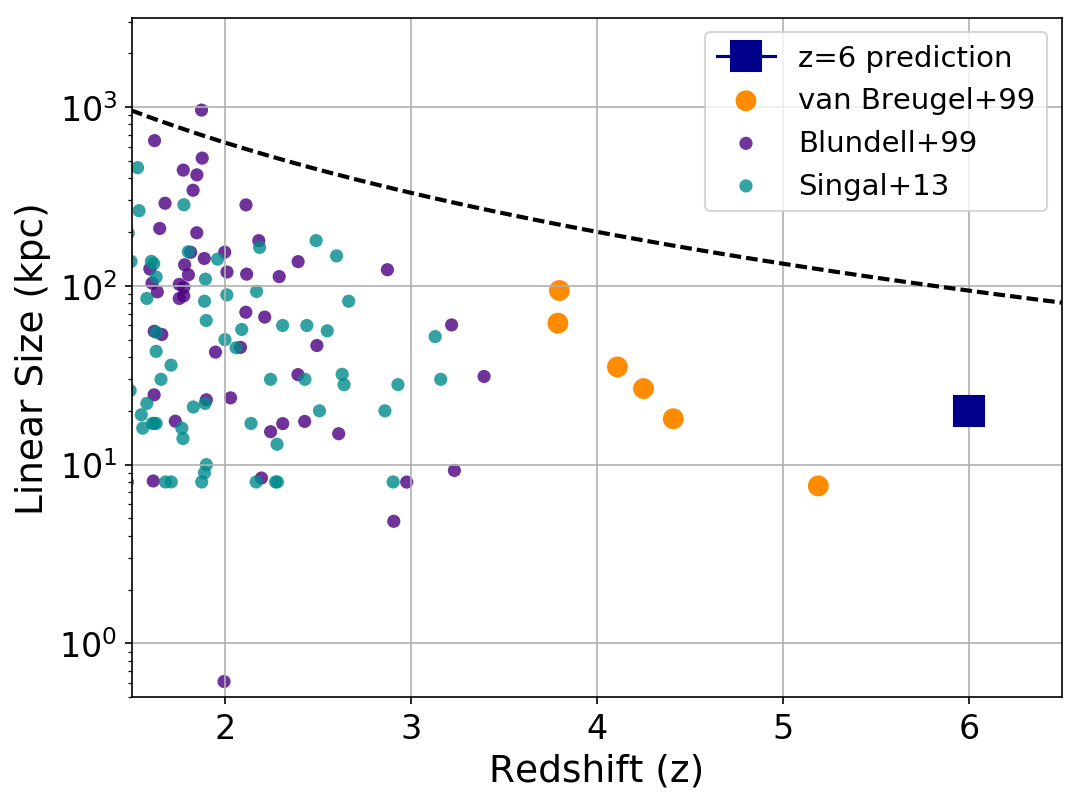}
	\caption{The median linear size of radio sources at $z=6$ with $S_{150} > 0.5$ mJy at 150 MHz in our simulation was found to be around 20 Kpc, shown as a blue square. Also shown are sizes of HzRGs taken from \citet{blu99} (purple points), \citet{sin13} (cyan points) and \citet{vbr99} (orange points), which have been updated using the cosmology used in this study. The dashed line shown for representative purposes is the evolution of the parameter $D_{IC}$, which we show to be a function of redshift (Equation 8). Our predictions seem to be in line with the generally decreasing trend of linear sizes with redshift.}
	\label{fig:z6_sizedist}
\end{figure}

Our size prediction at $z=6$ seems to be in line with the observed sizes of currently known high-redshift radio galaxies, as reported by \citet{blu99}, \citet{sin13}, and \citet{vbr99}, shown in Figure \ref{fig:z6_sizedist}. Also shown as reference is the redshift dependence of the parameter $D_{IC}$ (see Section 2.3.3) for a constant jet power (dashed line). It is worth noting that the samples used to compare our results are flux limited, with the \citet{blu99} sample containing sources with $S_{151} \ge 0.50$ Jy, \citet{sin13} sample containing sources with $S_{150} > 0.7$ Jy and sources in \citet{vbr99}, based on the sample of ultra-steep spectrum radio sources compiled by \citet{deb00}, having flux densities $S_{150} \ge 180$ mJy. These are likely to introduce various selection effects that may affect linear size determination that our model does not take into account. Additionally, the flux limits of the above mentioned samples may have implications on linear sizes being probed, as radio power and size seem to be correlated, as seen through the P-D tracks and as reported by \citet{ker12}. Regardless, the sizes predicted by our model seem to be in line with the observed trend.

\subsection{Comparison with exponential evolution of space densities}
For further analysis, we take into consideration the predicted $z=6$ radio luminosity function (RLF) at an average source age of $\sim0.32$ Myr. At this time step, the functional form of the RLF is well behaved and samples a broad range of luminosities. The resulting luminosity function is well fit with a double power law of the form
\begin{equation}
	\phi(P) = \phi_{\star} \times \left[\left(\frac{P}{P_{\star}}\right)^{\alpha} + \left(\frac{P}{P_{\star}}\right)^{\beta}\right]^{-1}
\end{equation}	
with parameters $\log \phi_\star = -9.05$ Mpc$^{-3}$ dex$^{-1}$, $\log P_\star = 27.91$ W Hz$^{-1}$, faint end slope $\alpha = 1.04$ and bright end slope $\beta = 2.92$.

We now compare the predicted RLF with a simple density evolution model for steep-spectrum radio sources obtained by evolving the relatively well constrained radio luminosity function at $z \sim 2$ determined by \citet{rig15}. Again, the space densities have been scaled to 150 MHz from 1.4 GHz at which they were originally calculated. The resulting space densities at 150 MHz are best fit with a double power-law, with parameters $\phi_{\star} = 1.80 \times 10^{-8}$ Mpc$^{-3}$ $(\log P_{150})^{-1}$, $\log P_{\star} = 28.95$ W Hz$^{-1}$, $\alpha = 0.72$ and $\beta = 2.74$. We then assume an exponential decline in space density with redshift, which can be written as $\phi(P_{150},z) = \phi(P_{150}, 2) 10^{q(z-2.0)}$, where $q$ is the evolutionary parameter. Several studies of quasars at various wavelengths have inferred the value of $q$ to range from $-0.59$ to $-0.43$ \citep{fan01, fon07, bru09, wil10, civ11, roc12}. Assuming the evolution of radio galaxies to match the evolution of optically-selected quasars at high redshifts, it seems reasonable to scale the radio luminosity function using an exponentially declining model that seems to work for quasars.

\begin{figure}
	\centering
	\includegraphics[scale=0.45]{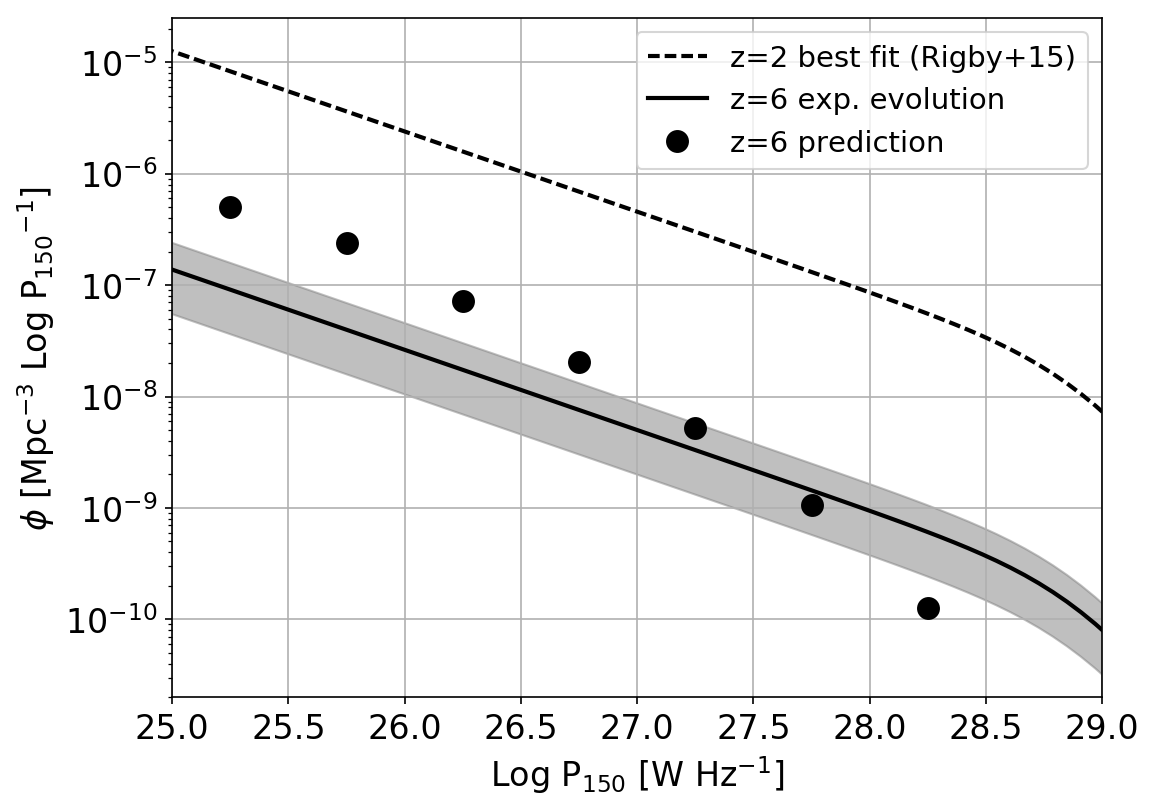}
	\caption{Comparison of the radio luminosity function at $z=6$ predicted by our model (black points) and the RLF obtained by using a pure density evolution model from $z=2$ (solid curve). The dashed curve shows the luminosity function at $z=2$ determined by \citet{rig15}. The evolutionary parameter found to best match our prediction at $z=6$ is $q=-0.49$, which indicates stronger evolution for radio galaxies but is within the observed evolution of optical and Xray quasar space densities at high redshifts. The shaded region shows RLFs expected with $-0.59<q<-0.43$.}
	\label{fig:lumfn_comparison}
\end{figure}

We find that $q=-0.49$ fits our predicted RLF best, which is shown in Figure \ref{fig:lumfn_comparison}. The dashed region represents $-0.59<q<-0.43$. The evolution we find is slightly stronger than what is observed in quasars. This is not surprising, as the dependence of radio luminosity on size and lifetime should introduce deviations from the evolution observed for quasars. Our model over-predicts space densities at the faintest end. This could be due to the fact that the faint-end slope of the black hole mass function we use as input is poorly constrained, which is most likely affecting the faint-end of our predicted RLF too. At the highest luminosities, space densities predicted by our model are lower than those expected from pure density evolution by a factor of 0.67 dex in the luminosity bin $\log P_{150} = 28.5-28.5$ and 2 dex in the bin $\log P_{150} = 28.5-29.0$, where our model barely predicts any sources. This is most likely due to the increasing dominance of inverse Compton losses at high redshifts, which lowers the space density of the brightest sources that grow and lose energy quicker. This effect is not accounted for by pure density evolution, suggesting that a luminosity-dependent density evolution (LDDE) model \citep{dun90, wil01, rig11, rig15} may be better suited to describing the evolution of space densities of radio sources out to high redshifts.  

\subsection{Implications of radio luminosity function at z = 6}
\label{sec:implications}
We now use the modelled RLF at $z=6$ to make predictions about number counts in current and future low-frequency radio surveys. Expected number counts are essential for designing surveys and observing strategies that target the identification of the highest-redshift radio galaxies.

\subsubsection{Number count predictions}
To calculate expected number counts, we integrate the RLF at $z=6$ down to flux limits chosen to represent various surveys at 150 MHz with instruments such as LOFAR\footnote{http://www.lofar.org/}, GMRT\footnote{http://www.gmrt.ncra.tifr.res.in/}, and MWA\footnote{http://www.mwatelescope.org/}, which are shown in Table \ref{tab:sourcecounts}. It is clear that the current and upcoming surveys with LOFAR will be the way forward to detecting a large number of $z\ge6$ sources. The preliminary LoTSS direction-independent (DI) survey covers roughly 350 sq. degrees \citep{shi17} and may lead to detection of around 32 radio galaxies at $z\ge6$. Direction-dependent (DD) calibration, which takes care of effects such as varying ionospheric conditions and errors in beam models, is currently ongoing on the LoTSS fields and will result in high-fidelity images at full resolution and sensitivity \citep{shi17}. On completion, the LoTSS direction-dependent survey shall provide a large sky coverage (Dec $> 0$), leading to potential detection of more than 12000 $z\ge6$ sources. The current highest-resolution survey at 150 MHz covering a very large area on the sky is the TGSS ADR, which covers around 37,000 sq. degrees above a declination of $-53$ \citep{int17}. In this survey, one could expect to detect around 92 sources at $z=6$. 

The probability of 21-cm absorption arising from the neutral intergalactic medium at $z>6$ depends on the optical depth of neutral hydrogen, $\tau$, that pervades the universe. This in turn depends on a number of key parameters such as the temperature of the CMB, spin temperature of neutral hydrogen and the neutral hydrogen fraction \citep{car02}. The detection of such features in the radio continuum of $z>6$ sources, however, depends on a number of instrumental properties too. \citet{cia13} investigated whether 21cm absorption can be detected by LOFAR and found that by using 48 stations and assuming the system temperature to be 200 K at an observing frequency of around 150 MHz, 21cm absorption features can be observed for a source at $z=7$ with a flux density of 50 mJy, along a line-of-sight with $\tau=0.12$. Such a detection would require an integration time of 1000 hours using a bandwidth of 5 kHz. Using a larger bandwidth may bring down the required integration time. Detection of a source with a flux density of 50 mJy at $z>6$, however, is very unlikely but there may be more than 30 sources with flux densities $>15$ mJy that could indeed be used for 21cm absorption studies \citep{car02}. The on-going direction-dependent all-sky survey with LOFAR will be extremely efficient in laying the groundwork for future 21cm absorption studies with the SKA, when much fainter sources can be used for detection of 21cm absorption. 

\begin{table*}
	\centering
	\caption{Predicted number of radio galaxies at $z = 6$ detected with $5\sigma$ in current and future surveys at 150 MHz.}

	\begin{threeparttable}
	\begin{tabular}{c c c c c c c}
		\hline
		Survey & Ref. & Flux lim (mJy/beam) & {N/deg$^2$} & Area (deg$^2$) & Total N\\ \hline \hline
		LOFAR Tier 3 (planned) & & 0.01 & 7.18 & & \\
		LOFAR Tier 2 (planned) & & 0.03 & 2.28 & & \\
		LoTSS$^a$ direction-dependent (ongoing) & \citet{shi17} & 0.1 & 0.63 & 20500 & 12915\\
		LoTSS direction-independent & \citet{shi17} & 0.5 & 0.092 & 350 & 32\\
		TGSS ADR & \citet{int17} & 3.5 & 0.0025 & 37000 & 92 \\
		GLEAM & \citet{way15} & 5 & 0.001 & 31000 & 31\\
		LOFAR MSSS HBA & \citet{hea15} & 10 & 0.0001 & 20500 & 2\\ \hline
	\end{tabular}
	\begin{tablenotes}
		\small
		\item $^a$ LOFAR Two-metre Sky Survey
	\end{tablenotes}
	\end{threeparttable}
	\label{tab:sourcecounts}
\end{table*}

\subsubsection{Observing strategies: Coverage area vs. depth}
\begin{figure}
	\centering
	\includegraphics[scale=0.45]{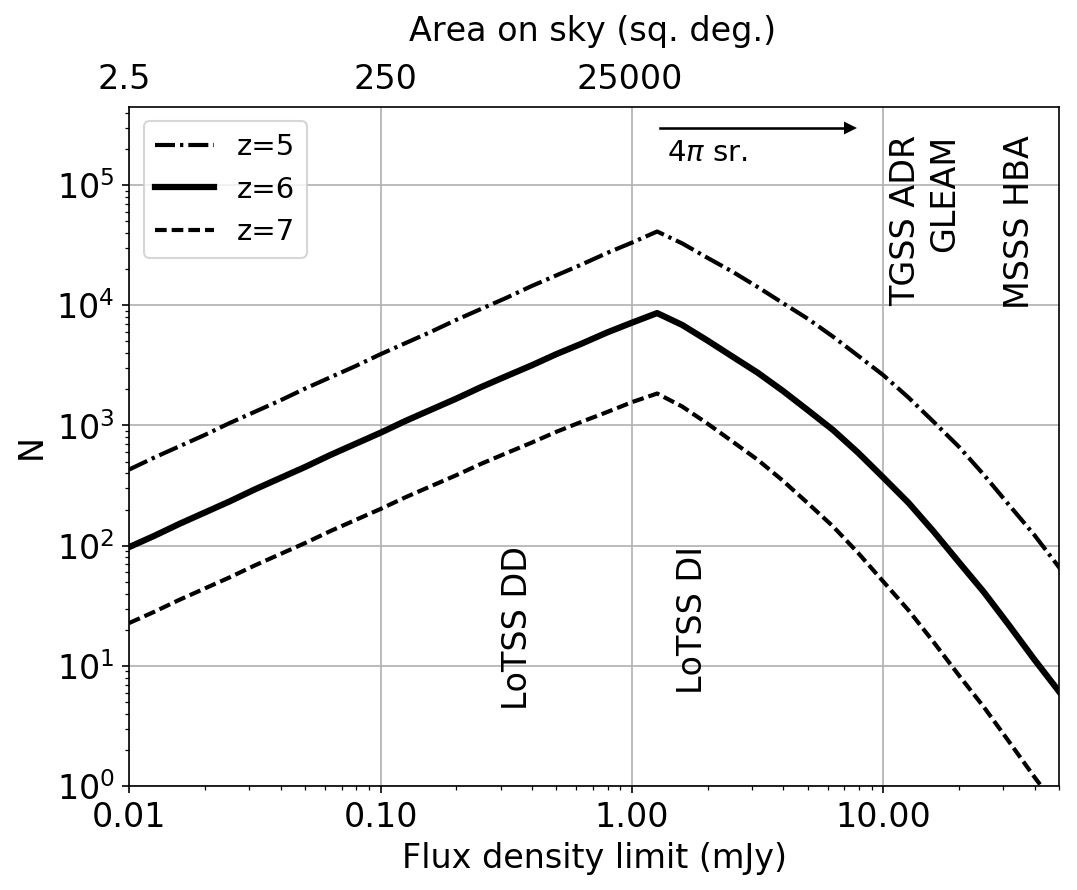}
	\caption{Number of high-$z$ radio sources expected to be detected as a function of flux density limits reached in a fixed observing time of 100 hours, using standard LOFAR configuration. The upper x-axis shows sky coverage in the given observing time at a particular flux density limit. Also shown are the $5\sigma$ limits of current and future low-frequency surveys. The LoTSS DI and DD surveys seem to be ideal to detect a large number of $z>5$ sources. This comparison further demonstrates that going too deep is not the best strategy as a large area of the sky needs to be surveyed to increase the number of detections.}
	\label{fig:areadepth}
\end{figure}

To compare what the ideal trade-off between depth and sky coverage would be that maximises the detection of $z\ge6$ sources in current and upcoming surveys, we calculate the total number of sources that could be detected in a set observing time. A few assumptions go into this calculation: a) we assume that it takes 8 hours to reach depths of 0.1 mJy and b) each `pointing' of the radio telescope covers 20 sq. degrees on the sky. These assumptions are based on the quoted values for LOFAR \citep{shi17}. 

For a fixed observing time of 100 hours, the expected number of sources for different noise levels achievable is shown in Figure \ref{fig:areadepth}. Quite clearly, the LoTSS DI and DD surveys would be ideal to detect sources at $z>5$. The luminosity functions at $z=5$ and $z=7$ are obtained by assuming pure density evolution with $q=-0.49$, which was found to best-fit our $z=6$ RLF prediction in the previous section. There clearly lies a `sweet-spot' in the coverage area vs. depth trade-off, which covers a large enough area and goes deep enough in a way that maximises detection of high-$z$ radio sources. Note that the $4\pi$ sr. limit is achieved just after a flux density limit of 1 mJy in the given observing time.

\section{Summary}
\label{sec:summary}
In this study, we have used a semi-analytical model based on prescriptions laid out by \citetalias{kai07} to predict the luminosity and linear size distribution of radio sources. The model takes as input the black hole mass function and Eddington ratio distribution at any epoch and implements simple energy loss mechanisms that dominate at different phases of a radio source's lifetime. Radio jet powers are assigned to each active black hole depending on the black hole mass and the Eddington ratio, which is randomly sampled from the input distribution. As the radio source grows, it initially loses energy predominantly by synchrotron emission when its magnetic field is strong. Adiabatic losses take over in the intermediate phase, with inverse Compton losses due to the CMB radiation dominating in the later stages of a source's lifetime.

We first implement our model at $z=2$ where sufficient data for radio luminosities and linear sizes is available. Making certain assumptions about the prevalent physical conditions and the black hole spin distribution, we predict a radio luminosity function that is consistent with observations. We are also able to reproduce the break in luminosity that marks the distinction between FRI and FRII radio source populations, supported by both theory and observations in the literature. We argue that this bi-modality in source population may be due to the accretion mechanism (thin disk vs. ADAF) that is responsible for powering the radio jets. Further, we are able to reproduce the distribution of linear sizes observed in flux limited surveys from the literature.

We then extend our modelling to $z=6$ where radio sources can be unique probes of the epoch of reionisation as 21cm absorption features in the radio continuum of a source at $z>6$ can be used to constrain the properties of the neutral inter-galactic medium in the very early universe. Using simplified assumptions about the black hole spin in the early universe, we predict a radio luminosity function at $z=6$. We show that radio sources at $z=6$ do not live for very long compared to typical ages of sources in the low-redshift universe. This is mainly due to radio jets being intrinsically weaker because of lower black hole masses and due to inverse Compton (IC) scattering being highly dominant in the early universe (due to a higher CMB energy density) that frustrates the jets and suppresses the growth of linear sizes of sources at high redshifts. Further, the predicted distribution of linear sizes is consistent with observations of the currently known highest redshift radio galaxies and the generally decreasing trend observed between linear sizes and redshift.

We compare the predicted $z=6$ RLF with a pure density evolution model from lower-$z$ for radio galaxies, based on the evolution of QSO luminosity functions. The evolution is of the form $\phi(P_{150},z) = \phi(P_{150}, 2) 10^{q(z-2.0)}$. We find that $q=-0.49$ fits reasonably well with luminosities between $10^{26.5} - 10^{28}$ W Hz$^{-1}$. There is some disagreement at the highest luminosity end and we attribute this to the significantly enhanced inverse Compton losses at higher redshifts, that have the most impact on luminous sources. Such an effect is not captured by a pure density evolution model and we argue that luminosity-dependent density evolution would better explain the redshift evolution of the radio luminosity function.

We finally predict the number of high redshift radio galaxies that may be observed in current and future low-frequency surveys with LOFAR, GMRT and MWA. To better understand the trade-off between coverage area and depth in a way that maximises detection of high-$z$ radio sources, we calculate the total number of sources that would be expected as a function of flux density limits for a fixed observing time. We show that the LOFAR Two-metre Sky Survey (LoTSS) direction-independent and direction-dependent calibration surveys sit at the sweet-spot of coverage area and depth, and should be most effective at detecting large numbers of $z>5$ radio sources. Detection of a 21cm absorption signal in the continuum of a radio source at $z>6$ will enable studies of the epoch of reionisation in unparalleled detail.

\section*{Acknowledgments}
We thank the referee for useful comments and suggestions. AS is grateful to Philip Best, George Miley and Kinwah Wu for fruitful discussions and comments over the course of this work. AS and HJR gratefully acknowledge support from the European Research Council under the European Union’s Seventh Framework Programme (FP/2007-2013)/ERC Advanced Grant NEWCLUSTERS-321271. EER acknowledges financial support from NWO (grant number: NWO-TOP LOFAR 614.001.006). 

This work has made extensive use of IPython \citep{per07}. This research made use of Astropy, a community-developed core Python package for Astronomy \citep{ast13}. The data analysis and visualisation were done using TOPCAT \citep{top05}. All figures used in this paper were produced using matplotlib \citep{hun07}. This work would not have been possible without the countless hours put in by members of the open-source community all around the world. 

\renewcommand{\bibname}{References}

\bibliographystyle{mnras}
\bibliography{bibliography}

\appendix

\section{Calculation of Inverse Compton loss distance}
\label{app:dic}
Here we show the calculation of the distance from the galaxy at which Inverse Compton (IC) losses begin to dominate. This is done by equating the predicted luminosities from the adiabatic loss phase and the IC loss phase. The expressions for luminosities are adapted from \citet{kai07}.

For a source in the adiabatic loss phase at an observing frequency of 150 MHz
\begin{equation}
	L_{150} = \frac{3f_L Q_{\textrm{jet}} p^{3/4}}{3 + a_1} t
\end{equation}
where $Q_{\textrm{jet}}$ is the jet power, $p$ is the pressure in the lobe and the constants $f_L = 3.4 \times 10^{-17} \left(\frac{0.15}{\textrm{GHz}}\right)^{-1/2}$ J$^{1/4}$ m$^{1/4}$ s$^{2}$ kg$^{-1}$, and $a_1 = 3/2$ \citep{kai07}. The pressure inside the lobe is given by
\begin{equation}
	p = f_p (\rho d^{\beta})^{1/3} Q_{\textrm{jet}}^{2/3} D^{(-4-\beta)/3}
\end{equation}

The luminosity of a source in the IC loss phase can be written as
\begin{equation}
	L_{150} = \frac{3 m_e c^2 f_n Q_{\textrm{jet}} p}{14 \sqrt{A} u^{\textrm{CMB}}_0 (1+z)^4 150 \textrm{ MHz}}
\end{equation}
where $m_e$ is the mass of an electron, $c$ is the speed of light, $u_{\textrm{CMB}}$ is the CMB energy density and the constants $f_n = 1.2 \times 10^{12}$ s$^{2}$ kg$^{-1}$ m$^{-2}$ and $A = 4$ \citep{kai07}. The present day CMB energy density, $u^{\textrm{CMB}}_0$ is $\sim 4 \times 10^{-14}$ J m$^{-3}$ and scales with redshift $z$ as $(1+z)^4$.

Linear size, $D$ is related to the age of the source, $t$ as 
\begin{equation}
	D \propto t^{3/(5-\beta)}
\end{equation}
For expansion into the ambient medium, we take $\beta = 2$ and therefore, there is a linear relation between source age $t$ and linear size $D$, $t \propto D$. We then equate equations A1 and A3 using A2 to calculate the distance at which IC losses would begin to dominate over adiabatic losses. This distance depends on the initial jet power of the source and is given by
\begin{equation}
	D_{IC} = \textrm{const} \times (\rho d^\beta)^{-1/6} Q_{\textrm{jet}}^{1/3} (1+z)^{-8/3}
\end{equation}
where the constant is given by
\begin{equation}
	\frac{m_e c^2 f_n f_p^{1/4} C (3+a_1)}{14 f_L \sqrt{A} u_{\textrm{CMB}}^0 150 \textrm{ MHz}}
\end{equation} 

The values of model parameters used in this study are shown in Table \ref{tab:params}.

\begin{table}
\centering
\caption{Fiducial model parameters used in this study, taken from \citetalias{kai07}.}
\begin{tabular}{c c}
	\hline
	Parameter & Value \\
	\hline \hline
	A & 4 \\
	$\rho$ & $10^{-22}$ kg m$^{-1}$ \\
	\multirow{2}{*}{$d$} &$2$ kpc at $z\le2$ \\
     & $2\times [(1+z)/3]^{-1.25}$ kpc at $z>2$ \\  
	$u_{\textrm{CMB}}^0$ & $4 \times 10^{-14}$ J m$^{-3}$ \\
	$f_n$ & $1.2 \times 10^{12}$ s$^2$ kg$^{-1}$ m$^{-2}$ \\
	$f_p$ & $0.11$ \\
	$C$ & $1.5$ \\
	$a_1$ & $3/2$ \\
	$f_L$ & $3.4 \times 10^{-17} \times 1/\sqrt{0.15}$ J$^{1/4}$ m$^{1/4}$ s$^2$ kg$^{-1}$ \\
	\hline
\end{tabular}
\label{tab:params}
\end{table}

% \bsp
\label{lastpage}
\end{document}